\documentclass[prl,noshowkeys,twocolumn,superscriptaddress]{revtex4}
\usepackage{graphicx}
\def\gap{\;\rlap{\lower 2.5pt
 \hbox{$\sim$}}\raise 1.5pt\hbox{$>$}\;}
\def\lap{\;\rlap{\lower 2.5pt
   \hbox{$\sim$}}\raise 1.5pt\hbox{$<$}\;}
\def\gsim{\;\rlap{\lower 2.5pt
 \hbox{$\sim$}}\raise 1.5pt\hbox{$>$}\;}
\def\lsim{\;\rlap{\lower 2.5pt
   \hbox{$\sim$}}\raise 1.5pt\hbox{$<$}\;}
\def\msun{{\rm\,M_\odot}}

\def\spose#1{\hbox to 0pt{#1\hss}}
\def\lta{\mathrel{\spose{\lower 3pt\hbox{$\mathchar''218$}}
     \raise 2.0pt\hbox{$\mathchar''13C$}}}
\def\gta{\mathrel{\spose{\lower 3pt\hbox{$\mathchar''218$}}
     \raise 2.0pt\hbox{$\mathchar''13E$}}}
\newcommand{\beq}{\begin{equation}}
\newcommand{\eeq}{\end{equation}}
\newcommand{\be}{\begin{equation}}
\newcommand{\ee}{\end{equation}}

\newcommand{\ls}{\mathrel{\raise1.16pt\hbox{$<$}\kern-7.0pt 
\lower3.06pt\hbox{{$\scriptstyle \sim$}}}}         
\newcommand{\gs}{\mathrel{\raise1.16pt\hbox{$>$}\kern-7.0pt 
\lower3.06pt\hbox{{$\scriptstyle \sim$}}}}         
\def\VEV#1{{\langle #1 \rangle}}
\long\def\comment#1{}

\def\mh{M_{\bullet}}
\def\msun{M_{\odot}}
\def\fun#1#2{\lower3.6pt\vbox{\baselineskip0pt\lineskip.9pt
  \ialign{$\mathsurround=0pt#1\hfil##\hfil$\crcr#2\crcr\sim\crcr}}}
\def\lap{\mathrel{\mathpalette\fun <}}
\def\gap{\mathrel{\mathpalette\fun >}}
\newcommand{\ba}{\begin{eqnarray}}
\newcommand{\ea}{\end{eqnarray}}

\begin{document}
\bibliographystyle{apsrev.bst}
\title{On Dark Matter Annihilation in the Local Group.}
\author{Lidia Pieri}
\email{pieri@fis.uniroma3.it}
\affiliation{Dept. of Physics, Universit\`a di Roma Tre, Rome, Italy}
\affiliation{INFN Roma Tre, Rome, Italy}
\author{Enzo Branchini}
\affiliation{Dept. of Physics, Universit\`a di Roma Tre, 
Rome, Italy}

\begin{abstract}
Under the hypothesis of a Dark Matter composed by supersymmetric 
particles like neutralinos, 
we investigate the possibility that their annihilation in the halos
of nearby galaxies could produce detectable fluxes 
of $\gamma$-photons. Expected fluxes depend on several, poorly known
quantities such as the density profiles of Dark Matter halos, the existence 
and prominence of central density cusps and 
the presence of a population of sub-halos.
We find that, for all reasonable choices of Dark Matter halo models,
the intensity of the $\gamma$-ray flux from some of the nearest extragalactic
objects, like M31, is comparable or higher than the diffuse Galactic 
foreground.
We show that next generation ground-based experiments could have 
the sensitivity to reveal such fluxes which could
help us unveiling the nature of Dark Matter particles.
\end{abstract}

\maketitle
\noindent

\section{I. Introduction}
The nature of the Dark Matter (DM) is a fundamental missing piece 
of the dark puzzle of the universe, and represents
one of the hottest challenges facing particle physics and cosmology today.
The latest observational data 
\cite{concordance} prefer a flat universe with 
a $\sim 23\%$ mass fraction of DM, seeded with Gaussian, scale dependent 
adiabatic perturbations.
The bulk of DM is believed to be ``cold'', 
i.e. composed by weakly interacting particles that were 
non-relativistic at the epoch of decoupling.
\noindent
The most popular candidate for Cold Dark Matter (CDM)
is the lightest supersymmetric 
particle (LSP) which, in most supersymmetry breaking scenarios, is 
the neutralino $\chi$. Neutralinos are spin-$\frac{1}{2}$ Majorana fermions, 
linear combination of the neutral gauge bosons and neutral
Higgs doublet spartners, 
$\chi^0 = a \tilde B + b \tilde{W_3} + c \tilde{H_1^0} + d \tilde{H_2^0}$.
In the popular SUGRA or SUGRA-like models, 
where gaugino-universality is required, 
its mass is constrained by accelerator
searches and by theoretical considerations of thermal freeze-out to lie
in the range $50 \ \mathrm{GeV} \lap m_{\chi} \lap 10 \ \mathrm{TeV}$
\cite{Ellis:00,Jungman:96}.
Neutralinos decouple at a temperature that is roughly $m_{\chi}/20$, 
hence they behave like CDM. 
If $R$-parity is conserved, neutralinos are stable and
can change their cosmological 
abundance only through annihilation,
which implies that the resulting production of detectable
matter, antimatter, neutrinos and
photons is enhanced in high density regions like the center
of virialized DM halos.
At low energies, direct detection terrestrial experiments are being performed
to measure the energy deposited by elastic neutralino-nucleon scattering
\cite{Nicolao:01,Feng:00a}.
Besides direct searches, indirect 
detection of DM through its annihilation products can be 
implemented \cite{Bergstrom:00,Feng:01}, which are necessary for energies 
greater than $250 \ GeV$. 
Most indirect searches rely on detection of 
antimatter \cite{bottino:98,donato:00}
and $\gamma$-rays produced in the Milky Way halo 
\cite{Urban:92,Berezinsky:94,Bergstrom:98} or 
in Galactic sub-structures \cite{Roldan:00,aloisio:02};
the possibility of looking at 
extragalactic sources has also been 
suggested \cite{Roldan:00,Ullio:02,Taylor:03,Bergstrom:01,Tyler:02,baltz:00,iro:03}. 
Unfortunately, our partial knowledge of those 
astrophysical processes that govern the assembly of DM 
concentrations did not allow to make firm predictions 
on DM detectability. 
In particular, the central structure of the DM halos
and the prominence of small scale structures within the
smooth DM halo are far from being well determined.
The purpose of this work is 
to explore the possibility of extending DM searches
to extragalactic sources in our Local Group 
of galaxies (LG) and to make robust predictions about
their actual detectability with ground-based experiments
once model uncertainties are taken into account.
Supersymmetric LSP are not the only particles which can produce
annihilation signals (e.g. \cite{Fargion:00}), yet in this work we focus
on the contribution of neutralinos only.

The outline of the paper is as follows: in section II we discuss the
theoretical framework and present the predictions from the smooth halo models. 
In section III and IV we show how predictions change when 
the presence of Supermassive Black Holes (SMBH) and that of a population
sub-galactic halos are taken into account. 
Section V deals with the actual detectability
of LG sources and Galactic sub-halos by 
ground-based detectors. Our conclusions are presented in Section VI.
\\

\section{II. Theoretical framework}

\noindent
In a phenomenological approach which considers 
the   LSP as a DM candidate, the expected photon flux from 
neutralino annihilation is given by: 

\begin{eqnarray}
\frac{d \Phi_\gamma (E)}{dE_\gamma} = 
\left [ \sum_{V = \gamma, Z} N_{V \gamma} b_{V \gamma} \delta 
\left ( E_\gamma - m_\chi(1 - \frac{m_V^2}{4 m_\chi^2}) \right ) + \right . 
 &  & \nonumber \\ 
 \left . + \sum_{F} \frac{d N_\gamma}{d E_\gamma} b_F \right ] 
   \frac{\VEV{\sigma_{a} v}}{2 m^2_\chi}  \int_{0}^{r_{max}}
 \frac{\rho_{\chi}^2 (r)}{{ D^2}}  r^2 dr \ \ .  & &
\label{flusso}
\end{eqnarray}

\noindent In this formula, the dependence of the flux
on particle physics inputs and on cosmological inputs
are factorized. \\

{\it Particle physics.} 
The first sum in brackets represents a $\gamma$-line,
i.e. the 1-loop mediated process $\chi\chi\rightarrow\gamma\gamma$ or  
$\chi\chi\rightarrow Z\gamma$. 
Since neutralinos move at Galactic speed \cite{Griest:00}, 
their annihilation occurs at rest and the outgoing photons carry an energy
equal ($\gamma \gamma$ final state) or very close ($Z \gamma$ final state) 
to the neutralino mass. For the corresponding 
$\gamma$-lines we have $N_{\gamma \gamma}=2$ and $N_{Z \gamma}=1$
monochromatic final state photons. Branching ratios 
$b_{Z \gamma} \sim b_{\gamma \gamma} \sim 10^{-3}$ imply 
that the photon flux is dominated by the continuum emission
rather than by the $\gamma$-lines.
The continuum emission is given by the second sum of Eq. (\ref{flusso}) 
running over all the tree-level final states.
At the tree-level neutralinos annihilate into fermions, gauge bosons, 
Higgs particles and gluons.
Decay and/or hadronization in $\pi^0$ give
a continuum spectrum of 
$\gamma$-photons emerging from the $\pi^0$ decay.
Depending on the annihilation channel, the continuum differential energy 
spectrum can be parametrized as $dN_\gamma/dx = a x^{-1.5} e^{-bx}$, 
where $x=E_\gamma / m_\chi$, $a \sim O(1)$ and $b\sim O(10)$. These 
spectra have been calculated in \cite{Bergstrom:98} 
using a PYTHIA Monte 
Carlo simulation.
\noindent Branching ratio $b_F$ calculations for each process 
involved in the sum and the evaluation of the total annihilation 
cross section $\VEV{\sigma_{a}v}$ 
depend on the assumed supersymmetric model. 
Here we consider as a unique 
annihilation channel the one into W bosons, $\chi\chi\rightarrow W^+ W^-$,
with parameters $a=0.73$ and $b=7.76$ in the above mentioned $dN_\gamma/dx$ 
formula.
We also use the constraint to the value of the thermally-averaged 
annihilation cross-section $\VEV{\sigma_{a}v}$ given by the neutralino
number density at the freeze-out epoch \cite{Jungman:96}: 
\begin{equation}
\Omega_\chi h^2 \sim \frac{3 \cdot 10^{-27} cm^3 s^{-1}}{{\VEV{\sigma_{a}v}}} ,
\end{equation}
where $h$ is the Hubble constant, $H$, 
in units of $100 \ km \ s^{-1} \ Mpc^{-1}$
and $\Omega_\chi$ is the average density of neutralinos in units of critical 
density, $\rho_c = 3 H^2 / 8 \pi G$.

\noindent Unless otherwise specified, throughout the paper we consider 
neutralinos with a mass of  $1 \ TeV$ and  
cross section $\VEV{\sigma_{a} v} = 2 \cdot 10^{-26} cm^{-3} s^{-1}$. \\

{\it Cosmology.} 
The last term in  Eq. (\ref{flusso}) describes the geometry of the
problem and assumes that the DM 
is concentrated in a single, spherical DM halo of radius $r_{max}$  
and density profile $\rho_\chi(r)$ located at distance $D$ from the observer.
The integral can be easily generalized to any DM distribution once that
the DM density 
along the line of sight, $\rho_\chi(r)$, is specified.

An accurate evaluation of this integral should account for the fact that
DM halos are triaxial rather than spherical objects
\cite{Moore:01}. Deviations from sphericity do not 
change appreciably the $\gamma$-ray flux from LG objects
within the typical acceptance angle of the detectors 
($0.1^{\circ}-1^{\circ}$) and would only enhance the 
Galactic flux by a modest $15 \%$ 
 \cite{stoher:03} and thus will not be considered here.
The shapes of the halo density profiles are poorly constrained by observations,
especially near the center. Indeed, recent measurements of the
dwarf galaxy rotation curves, as well as of the intracluster medium and of the
gravitational lensing,  
could not discriminate between dark halos with constant density core
and $r^{-1}$ cusps \cite{vandenbosh:01}.
Similarly, analytical arguments did not yet lead to a definitive,
unique prediction \cite{nusser:99,subramanian:00,dekel:02}.

The most stringent constraints on the shape of DM halos 
are currently obtained from N-body simulations.
Although the resolution in numerical experiments is still an issue
\cite{Power:03}, advances in numerical techniques have 
allowed to discover that DM halos have  universal density 
shapes $\rho_\chi(r) \sim \rho_s (r/r_s)^{-\alpha}$.
While singular isothermal halos characterized by a steep
$r^{-2}$ cusp are ruled out, current numerical experiments
do not allow to discriminate among the Moore profile 
\cite{Moore:99} characterized by $\alpha = 1.5$ and the shallower
NFW profile \cite{Navarro:97} with $\alpha = 1$ (or slightly
shallower ones \cite{stoher:03}).
In this paper we assume that the actual shape of the DM 
density profiles 
is bracketed by the Moore and the NFW models and thus we can 
obtain a fair estimate of current model uncertainties
by considering both of them in our analysis.

\noindent
Both models have 
central cusps where the neutralino annihilation rate is greatly enhanced.
The scale radius, $r_s$,  and the scale density, $\rho_s$, can be fixed 
by observations (the virial mass of the halo or its rotation velocity) 
and theoretical considerations that allow to determine 
the concentration parameter $c = r_{vir}/r_s$ (where the virial 
radius, $r_{vir}$, is defined as the radius within which the halo 
average density is $200  \rho_c$).

\noindent
Since the largest $\gamma$-ray fluxes are expected from nearby
objects, we have considered only the 44 nearest LG galaxies and their
parent halos. Table \ref{tab1} lists the 
masses, positions (from \cite{Mateo:98}) and virial radii 
of the galaxies which are more relevant for our analysis,
along with their scale radii and scale densities 
computed assuming a Moore profile.
Their concentration parameters, 
$c_{NFW}$ and  $c_{Moore}=0.64 \ c_{NFW}$, have been computed  
according to \cite{ENS} assuming
CDM power spectrum with a shape parameter $\Gamma = 0.2$,
normalized to $\sigma_8 = 0.9$.
In our analysis we have also included the
giant elliptical galaxy M87 at the center of the Virgo cluster
whose predicted annihilation flux was already computed by  \cite{baltz:00}.
The mass, distance and the virial radius 
of M87, listed in Table \ref{tab1}, are taken from \cite{Mclaughlin:99}. \\
\noindent 
Since the intensity of the $\gamma$-ray flux
is proportional to $\rho_{\chi}^2$, 
great attention must be paid in modeling
the central region of the halo. 
The fact that the Moore model predicts
a divergent flux from the halo center implies that there must exist 
a minimum radius, $r_{min}$, within which the self-annihilation rate
$t_{l} \sim (\VEV{\sigma_{a} v} \ n_\chi (r_{min}))^{-1}$ 
equals the dynamical time 
$t_{dyn} \sim (G \bar{\rho_{\chi}} )^{- \frac{1}{2}}$,
where $\bar{\rho_{\chi}}$ is the mean halo density
within  $r_{min}$ \cite{Berezinsky:92}.
We have used this prescription to
determine $r_{min}$ for all our 44 halos and model
their DM distribution with a NFW or a Moore profile 
with a small constant core $ \rho_\chi (r_{min})$.
The resulting profiles are
\begin{eqnarray}
\rho_\chi(r) & = & \frac{\rho_s^{nfw}}{\left( \frac{r}{r_s^{nfw}} 
\right) \left(1 + \frac{r}{r_s^{nfw}} \right)^2 }, 
\ \ \ r > r_{min} \nonumber  \\ \nonumber
\rho_\chi(r) & = & \rho_\chi^{nfw}(r_{min}), \ \ \ r \leq r_{min} \\
\end{eqnarray}
for the NFW case, and 
\begin{eqnarray}
\rho_\chi(r) & = & \frac{\rho_s^{moore}}{\left( \frac{r}{r_s^{moore}} 
\right)^{1.5} \left[1 + \left (\frac{r}{r_s^{moore}} \right)^{1.5} 
\right] }, \ \ \ r > r_{min} \ \ \ \nonumber \\ \nonumber
\rho_\chi(r) & = & \rho_\chi^{moore}(r_{min}), \ \ \ r \leq r_{min} \\
\end{eqnarray}
for the Moore case. 

\begin{figure} 
\includegraphics[height=6cm,width=8cm]{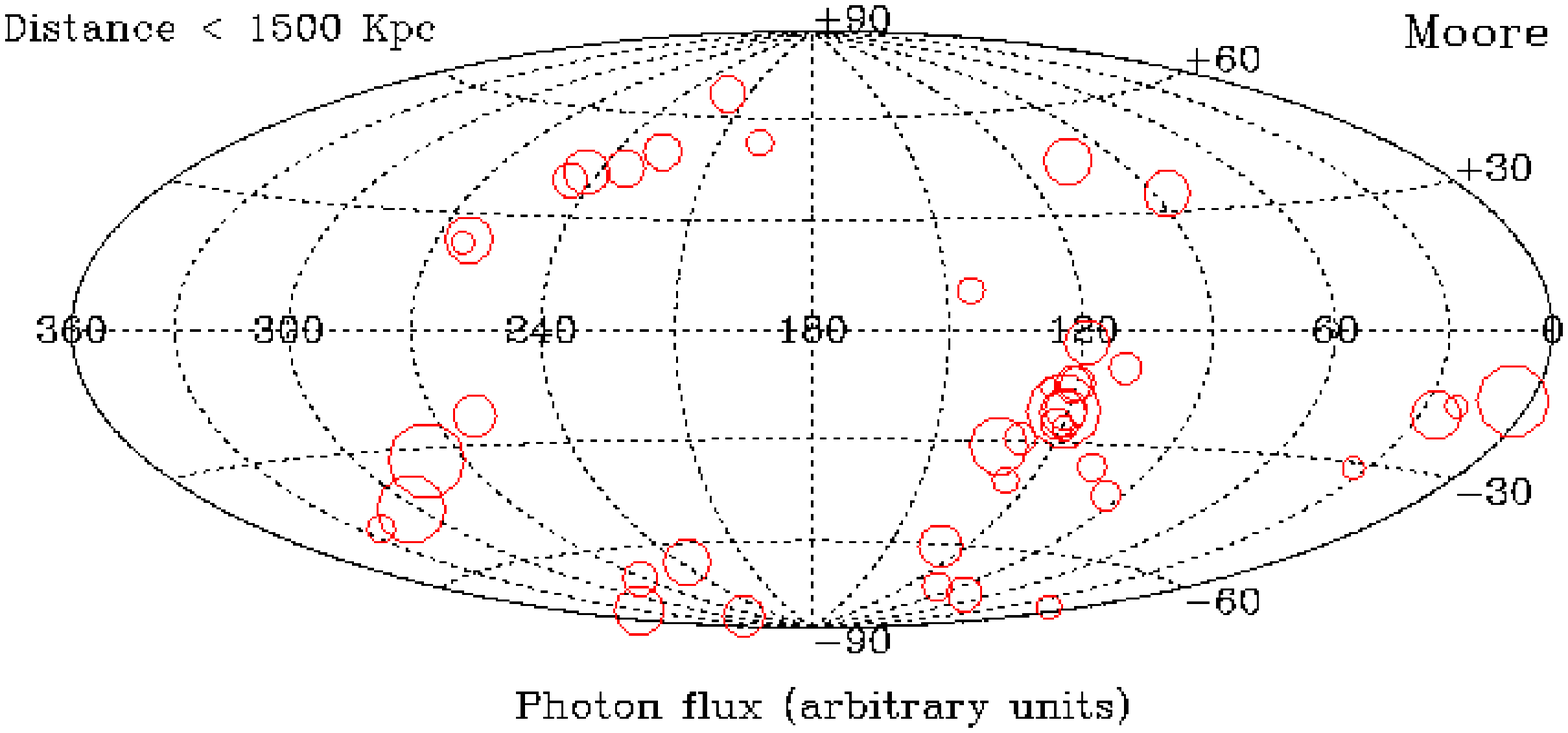}
\caption{Aitoff projection of the 44 nearest LG galaxies
in Galactic coordinates.
The size of each symbol is scaled to the $\gamma$-ray flux emitted
by a host DM halo with a Moore profile  
within a viewing angle of $1^{\circ}$ from the halo center. 
}
\label{fig1}
\end{figure}

Fig. \ref{fig1} shows an Aitoff projection of the angular position
of these LG galaxies in Galactic coordinates.
The size of each point is proportional to the 
$\gamma$-ray flux from Moore models, measured within a viewing angle of
$1^{\circ}$. 

\noindent
As shown by \cite{Roldan:00},
high energy $\gamma$-ray flux from nearby extragalactic halos
with a Moore density profile can be larger than the 
$\gamma$-ray foreground produced by neutralino annihilation in our Galaxy.
This is not an entirely trivial point as distant extragalactic 
objects are not resolved within the typical  $1^{\circ}$
angular resolution of the $\gamma$-rays detectors.
To check whether our conclusions still hold for shallower 
density profiles,
we have computed the $\gamma$-ray flux using both Moore and NFW
models for our LG halos
and compared it to the Galactic foreground.
In this work all fluxes are computed above $100 \ GeV$
within an angle $\Delta \Omega=10^{-3} sr$.
Note that none of the particle physics assumptions affect 
the ratio between Galactic and extragalactic fluxes that
we will focus on.

\begin{figure} 
\includegraphics[height=7cm,width=8cm]{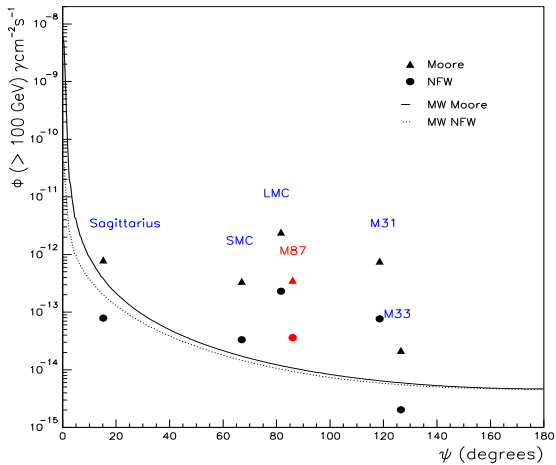}
\caption{Integrated $\gamma$-ray flux within $\Delta \Omega=10^{-3} sr$
as a function of the angular distance from the GC.
Filled triangles and filled dots represent the 
fluxes from brightest extragalactic objects predicted assuming a Moore
and a NFW profile, respectively. The continuous and the dashed curves
show the predicted Galactic foreground for the two model density profiles.}
\label{fig2}
\end{figure}

Fig. \ref{fig2} shows the smooth $\gamma$-ray
Galactic foreground for a Moore (continuous line) and a NFW (dotted line)
profile as a function of $\psi$, the angle from the Galactic Center (GC).
Filled triangles represent the $\gamma$-fluxes 
from those LG galaxies which are brighter than the Galactic 
foreground and from M87, assuming a 
Moore profile. 
The fluxes from the Small and Large  
Magellanic Clouds (SMC, LMC), M31 and, to a 
lesser extent, M87
are above the Galactic level.
In some cases, like M33 and Sagittarius, 
extragalactic and Galactic contributions
are comparable. 
Similar considerations are valid for 
the case of a NFW profile (filled dots),
despite of the significant reduction of 
the extragalactic fluxes that cause 
M33 and Sagittarius to be dimmer than the 
Galactic foreground.

\begin{table}[h]
\begin{center}
\begin{tabular}{|c|c|c|c|c|c|}
\hline \hline
Galaxy & mass & distance & $r_{vir}$ & $r_s^{moore}$ & $\rho_s^{moore}$ \\
name & ($\msun$) & ($kpc$) & ($kpc$) & ($kpc$) & ($\msun kpc^{-3}$) \\ \hline \hline
MW & $1.0 \cdot 10^{12}$ & 8.5 & 205 & 34.5 & $1.1 \cdot 10^6$ \\ \hline
Sagittarius & $9.4 \cdot 10^8$ & 24 & 20 & 2.5 & $2.3 \cdot 10^6$  \\ 
SMC & $2.5 \cdot 10^9$ & 58 &  28 & 3.6 & $2.1 \cdot 10^6$ \\ 
LMC & $1.4 \cdot 10^{10}$ & 49 & 49 & 6.8 & $1.8 \cdot 10^6$ \\ 
M31 & $2.0 \cdot 10^{12}$  & 770 &  258 & 47.3 & $8.6 \cdot 10^5$ \\ 
M33 & $4.0 \cdot 10^{10}$ & 840 & 70 & 10.6 & $1.6 \cdot 10^6$ \\ \hline
M87 & $4.2 \cdot 10^{14}$ & $1.5 \cdot 10^4$ & $1.55 \cdot 10^3$ & 461 & $2.6 \cdot 10^5$ \\ \hline
\hline

\end{tabular}
\caption{\label{tab1} Input parameters for the $\gamma$-ray 
flux prediction are shown for 
our Galaxy, M87 and for the LG galaxies brighter than the Galactic 
foreground. Masses and distances are taken from
\cite{Mateo:98}. Virial radii were calculated following the prescription 
given in the text. Scale radii and scale densities
were computed assuming a Moore profile.}
\end{center}
\end{table}

\section{III. Black holes, cusps and cores}

The previous results do not account for
those astrophysical processes which 
may modify the density near the halo center,
like the presence of a central SMBH.
The  effect of  a SMBH on the central shape
of a DM profile is, in fact, rather 
controversial.
The adiabatic growth of a SMBH of mass $\mh$ at the center of a DM
halo would steep the slope of the density profile to $\rho_{halo} \propto
r^{-\gamma}$, $2.2<\gamma<2.5$ 
within a radius $r_h \equiv G\mh/\sigma_v^2$, where 
$\sigma_v$ is the 1-D {\it RMS} velocity dispersion of the DM particles 
\cite{Gondolo:99}.
As a consequence, the expected $\gamma$-ray emission would be larger in
galaxies hosting SMBHs. 
However, when one takes into account dynamical processes such as 
merger and orbital decay of an off-centered SMBH \cite{ullio:01},
this central `spike' disappears or is greatly weakened.
Moreover, in  a CDM cosmology the assembly of black holes at the center of
galaxies is the end-product of the hierarchical build up of their parent halos.
In this scenario pregalactic black holes become incorporated
through series of mergers into larger and larger halos, sink to the center
owing to dynamical friction, form a binary system   \cite{Begelman:80} 
and eventually coalesce \cite{Peters:64} with 
the emission of gravitational waves.
The process of formation and decay of black hole binaries 
transfers angular momentum to the DM particles,
ejects them  via gravitational slingshot effect
and thus decreases the mass density 
around the binary \cite{Merritt:01,Milos:01}.
The net result is either a constant density core 
\cite{volonteri:03} or a shallower DM density profile 
$\rho_\chi \sim r^{-0.5}$ within a radius $r_{cut}= 10 - 100 \ pc$,
with a consequent decrease of the neutralinos annihilation 
rate \cite{Merritt:02}. 

\begin{figure} 
\includegraphics[height=7cm,width=8cm]{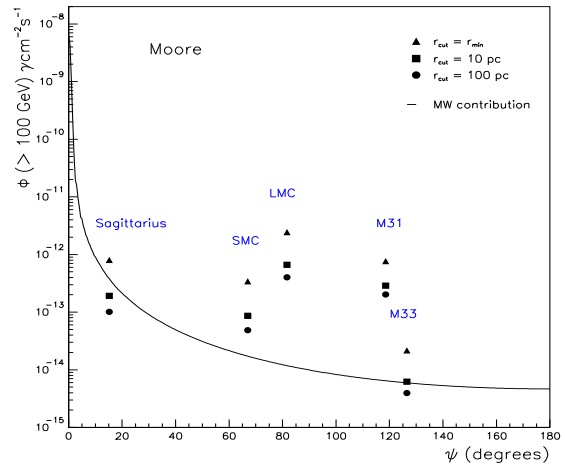}
\caption{Integrated $\gamma$-ray flux from the brightest 
LG members computed assuming a standard Moore profile
(filled triangles) and a Moore profile truncated at 10 pc and 100 pc
(filled squares and filled dots, respectively). The continuous
line show the expected Galactic annihilation foreground.}
\label{fig3}
\end{figure}

Observations indicate that SMBHs are only found in large galaxies.
Indeed, among the LG galaxies only the Milky Way and M31 harbor a SMBH, 
while a possible black hole in M33 would have a mass smaller than 
$3000 \msun$ \cite{Merrittf:01}. Assuming that the presence of a
SMBH creates a constant density core of similar size in both 
the dark and stellar components, we expect
a small core $r_{cut}= 0.38 \ pc$ in our Galaxy \cite{Genzel:00}
and one of  $r_{cut} \sim 10 \ pc$ at the center of M31 \cite{volonteri:03}.
As a consequence, the SMBH phenomenon would only decrease the
expected $\gamma$-ray
flux from M31 by a factor of 3, still well above the Galactic foreground.
In fact, our predictions do not change much 
even under the somewhat extreme hypothesis that all DM halos 
hosting LG galaxies have a constant density core of size
$10-100 \ pc$. Indeed, when repeating our calculation 
using both NFW and Moore profiles truncated 
at $10$ and $100 \ pc$ we do find that
the DM depletion near the halo center causes a significant decrease
of the $\gamma$-ray flux which, however, does not prevent
a few objects to shine prominently above the Galactic foreground.
Results are shown in Fig. \ref{fig3} in which the fluxes
from Moore DM halos (filled triangles), already shown in Fig. \ref{fig2},
are compared to the cases of Moore profiles
truncated at $r_{cut}= 10 \ pc$
(filled squares) and at $r_{cut}= 100 \ pc$ (filled circles).

\section{IV. Halo sub-structures}

Despite its remarkable success in explaining the large majority of
current observations, the excess 
power on small scales in the 
$\Lambda$CDM concordance model  
has lead to what is sometimes called {\sl the CDM crisis}. 
One aspect of this problem is the 
presence of a density spike at the center of DM halos, that is
difficult to reconcile with the observed rotation curves of nearby 
dwarf galaxies \cite{vandenbosh:01}. A second aspect is related to
the measured zero-point of the Tully-Fisher relation for spiral galaxies 
\cite{Tully:77}, which is smaller than the CDM-predicted one.
Whether these discrepancies are genuine or artifacts arising from observational
biases (e.g. beam-smearing effect in determining the rotation velocities
of dwarf galaxies) or numerical effects (e.g. limited resolution in 
CDM N-body experiments) is still matter of debate.
A second problem, also related to 
the excess small-scale power in the CDM model, is the
wealth  of sub-galactic, virialized halos that
do not appear to have an observational counterpart 
(the so called {\sl satellite catastrophe},
\cite{Moore:99}).
Numerical simulations have indeed shown that a CDM halo of
the size of our Galaxy should have about 50 satellites with circular velocities
$v_{circ}> 20 \ km s^{-1}$ instead of the dozen observed ones \cite{Klipin:99}.
Due to the hierarchical nature of the clustering process in a CDM
scenario, the mismatch is also found on larger systems of the size of the LG.
Model predictions for  halos with $ v_{circ} > 35 \ km s^{-1}$ might
be reconciled with observations by delaying the formation epoch of DM halos or by
computing their peak rotation velocity with different techniques \cite{Hayashi:02}.
For lighter halos, however, the lack of a luminous counterparts could 
be either ascribed to astrophysical processes (supernovae explosion,
stellar feedback, gas heating by ionizing radiation of cosmic origin)
that suppress star formation  or to a genuine disruption of the DM sub-halos 
by gravitational tides within their massive host. 
Finally, invoking Warm Dark Matter scenario decreases the small scale power 
and thus avoid the satellite catastrophe but would postpone the 
reionization of the universe to an epoch too late to be consistent with the recent
WMAP observations of the cosmic microwave background \cite{Yoshida:03}.\\

Here we account  for the presence of a 
population of DM sub-halos and investigate their 
effect on the neutralino annihilation signal.
Model predictions and related uncertainties are 
obtained using both numerical experiments and
thoretical arguments.

Large N-body simulations \cite{Moore:99,Ghigna:99}
have shown that DM halos host a population of sub-halos
characterized by a distribution function giving the probability of finding
a sub-halo of mass $m$ at distance $r$ from the halo center, that can 
be modeled as follows:
\cite{Blasi:00}:
\begin{equation}
n_{sh}(r,m) = A \left (\frac{m}{m_{H}} \right )^{-1.9} \left [1+ \left (\frac{r}{r^{sh}_c } \right )^2 \right]^{-1.5}
\label{distr}
\end{equation}
where ${r^{sh}_c}$ is the core radius of the sub-halos distribution,
$m_{H}$  is the mass of the parent halo and $A$ is a constant set 
to obtain 500 sub-halos of mass $ \ge 10^8 \msun$ in a halo with 
$m_{H} =2 \cdot 10^{12} \msun$ \cite{Ghigna:99} .
\noindent
The annihilation rate within sub-halos may contribute significantly 
to the total $\gamma$-ray flux. 
According to
\cite{Roldan:00,aloisio:02} 
the  $\gamma$-ray Galactic foreground
can be enhanced by over two orders of magnitude by the presence of a 
population of
sub-halos, an effect that may also help to explain the 
$\gamma$-ray flux excess found in EGRET data \cite{Bergstrom2:98}. 
To account for this effect, we have populated all the halos of our LG galaxies
including our own with a population of sub-halos according to the 
distribution function
(\ref{distr}) and computed their contribution to the $\gamma$-ray 
annihilation flux.
For each sub-halo we use the same density profiles 
as its massive host.
Current N-body simulations do not allow to model Eq. (\ref{distr})
very precisely. 
In particular, the mass clumped in sub-halos, $m_{cl}$,
and the value of ${r^{sh}_c}$ are poorly constrained. 
To account for these model uncertainties we have
generated several different sub-halos distributions that explore the edges 
of the  parameter space $[m_{cl}, {r^{sh}_c}]$. 
Following \cite{Blasi:00} we  have set $m_{cl}$ equal to  10 \% and 20 \%
of the total DM halo's mass, $m_H$, and the 
core radius, ${r^{sh}_c}$, to be 
5 \% and 30 \% of the parent halo virial radius, $r^H_{vir}$.
Sub-halos were generated in a mass range $[m_{min},0.1 \ m_H]$
and within a distance $r^H_{vir}$ from the halo center.
We set $m_{min}=10^6 \msun$  for Galactic halos and $m_{min}=10^7 \msun$
for the the extragalactic ones since the slope of the sub-halo mass 
distribution, Eq. (\ref{distr}), is shallow enough to ensure 
that for a fixed value of $m_{cl}$ 
the contribution  to the total $\gamma$-ray flux from halos with masses
smaller than  $m_{min}$
becomes negligible  \cite{aloisio:02,stoher:03}.

Very high resolutions N-body experiments show
that in a CDM scenario a large fraction of
the sub-halo population 
survive the complete tidal disruption \cite{Moore:01}.
However, these sub-halos are
tidally stripped of a fraction of their mass 
originating debris streams
and possibly changing their original profile 
at all radii, thus lowering 
the annihilation luminosity of the accreted systems
 \cite{Hayashi:02,helmi:03,stoher:03}.
Because of their finite resolution, the N-body experiments 
used to model Eq. (\ref{distr}) underestimate the 
role of these tidal fields and therefore do not allow
a reliable sampling of the low mass tail of the sub-halo distribution 
at small $r$. 
To better account for gravitational tides we follow \cite{Hayashi:02} and 
 use the ``tidal approximation'', i.e. 
we assume that all the mass beyond the sub-halo tidal radius, 
$ r_{tid}$, is lost in a single orbit without affecting its central 
density profile. The tidal radius is defined as the distance from 
the sub-halo center at which 
the tidal forces of the host potential equal the self-gravity of the sub-halo.
In the Roche limit:  
\begin{equation}
r_{tid} (r) =  \left (\frac{m}{2m_H(<r)} \right)^\frac{1}{3} r,
\end{equation}
where  $r$ is the distance from the halo center.
We truncate all sub-halos at 
their tidal radii and discard those with $ r_{tid}<r_s$, 
for which the binding energy is positive 
and the system is dispersed by tides. 
Even in this case, we estimate the model uncertainties by considering three
different cases: no tidal stripping,
tidal radius computed at the mean orbital radius and 
tidal radius computed at the pericenter of the orbit
(which is typically 1/5 of the mean orbital radius).

\begin{figure} 
\includegraphics[height=7cm,width=8cm]{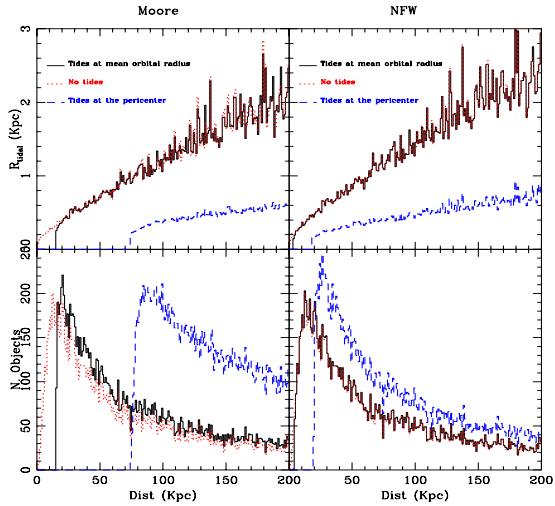}
\caption{Number of Galactic halos (bottom) and their 
average tidal radius (top) as a function of the
Galactic radius for a  Moore (left) and NFW (right)
density profile. Short-dashed histogram: no tidal stripping allowed.
Continuous line:
 tidal radius computed at the mean orbital radius.
Long dashed: tidal radius computed at the pericenter of the 
sub halo-orbit. In all cases shown $m_{cl}=20 \% \ m_H$ and 
 $r^{sh}_c = 0.05 \ r^H_{vir}$.
}
\label{fig4}
\end{figure}

The short-dashed histograms in the
 two bottom panels of Fig. \ref{fig4}
show the radial distribution of all
Galactic sub-halos with $M>10^6 \  \msun$,
distributed according to Eq. (\ref{distr}) 
in which $m_{cl}=20 \% \ m_H$ and 
$r^{sh}_c=5 \% \ r^H_{vir}$ for both a Moore
(left) and a NFW (right) density profile.
Including the tidal stripping effect 
destroys all sub-halos near the Galactic Center 
and modifies their radial distribution.
The effect is larger for the Moore model (which
is less concentrated than the NFW profile) and increases
with the strength of the tidal field 
(i.e. when tides are computed at the pericenter
of the orbit, long-dashed histogram).
The two top panels show the average tidal radius, 
$r_{tid}$ of the sub-halo population that survive tidal stripping.

\noindent
The effect of sub-halos on the emitted annihilation flux
is shown in Fig. 5a
for the case of a Moore profile and in Fig. 5b for a NFW profile.
The histogram drawn with a dotted line shows the diffuse $\gamma$-ray
flux obtained by adding the contribution to the smooth Galactic emission
of a sub-halo population distributed according to Eq. (\ref{distr}) in which 
$m_{cl}= 20 \% \ m_H$ and $r^{sh}_c= 5\%\ r^H_{vir}$.
Filled dots show the flux from the most prominent external objects,
computed by  adding together the halo flux to that of all 
their sub-halos within $1^{\circ}$ from the center
and by neglecting gravitational tides.
The presence of sub-halos dramatically boosts up the level
of the Galactic foreground (dot-dashed curve) 
by a factor 10-100, depending on the halo profile,
while increasing the flux from extragalactic sources 
by a more modest factor 2-5. 
As a consequence, the Galactic foreground 
becomes dominated by the local sub-halo emission,
resulting in the irregular
$\gamma$-ray Galactic brightness profiles
that outshine all extragalactic objects but M31 and the LMC.

These results, that are consistent with those of \cite{Roldan:00,aloisio:02}, 
change significantly when tidal fields are accounted for.
The net effect, that is more apparent for a Moore density 
profile, is that of  decreasing 
significantly the Galactic
$\gamma$-ray foreground, as shown by the histogram drawn with 
a continuous line in Figs. 5a and 5b. 
The flux from extragalactic objects 
(filled squares) is comparatively less affected 
by tidal effects and in some cases may actually 
be enhanced when the chance 
of bright foreground extragalactic sub-halos
becomes non-negligible.
As a result, taking into account the tidal stripping effects increases the 
prominence of extragalactic sources over the Galactic foreground.
If one further accounts for the eccentricity of the sub-halo
orbits and assume tidal stripping occur at 
their pericenter then the  
Galactic $\gamma$-ray foreground (dashed histogram) 
and the flux from extragalactic objects (filled triangles) further
decrease to a level similar to that of Fig. \ref{fig2}.

\begin{figure} 
\includegraphics[height=7cm,width=8cm]{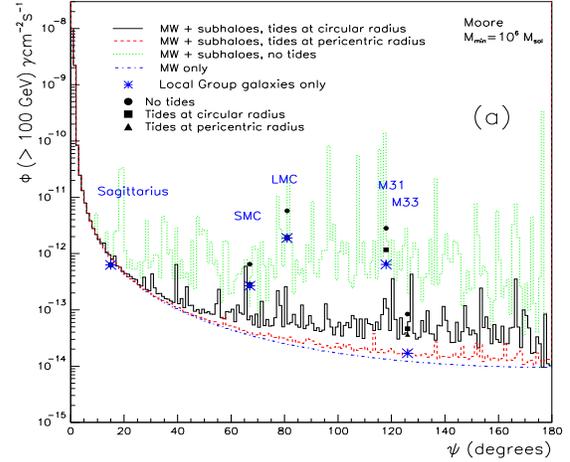}
\includegraphics[height=7cm,width=8cm]{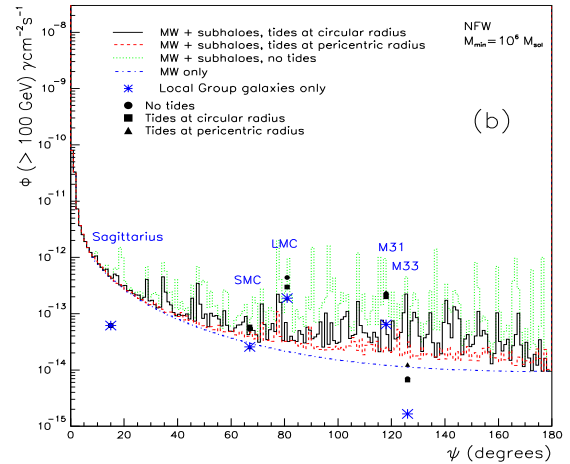}
\caption{Integrated $\gamma$-ray flux within 
$\Delta \Omega=10^{-3} sr$ for Moore (top) and
NFW (bottom) DM halos populated with sub-halos.
Galactic (histograms) and extragalactic 
(symbols) fluxes  are shown for the different sub-halo
populations described in the text.}
\label{fig5}
\end{figure}

To better quantify the uncertainties in modeling the sub-halo
population  we list in 
Table \ref{tab2} the fractional variation of the  
$\gamma$-ray flux from the extragalactic sources when 
varying the model parameters. The quantity displayed is
$\delta \Phi=(\Phi^{sh}-\Phi^{0})/\Phi^{0}$,
where $\Phi^{0}$ is the flux from a smooth profile while the flux
$\Phi^{sh}$ also accounts for the sub-halo population.
We have focused on the two brightest LG objects, M31 and the LMC. 
It is worth noticing that, due to their different distances from 
our Galaxy, while the flux from M31 is produced in the central
16-36 Kpc (depending on the model profile), the LMC flux is
produced in a much smaller region of radius 1-1.4 Kpc.
The various models explored are listed in Table \ref{tab2} 
and are characterized by a label in 
which the first two digits indicate the extent of 
the core radius of the sub-halo distribution (e.g. 30
means  ${r^{sh}_c} = 30\%  \ r^H_{vir}$), the subsequent two
digits characterize the mass clumped in sub-halos 
(e.g. 20 for $m_{cl} = 20\% \  m_H$) and the last 
two letters indicates the model tidal field  ($d0=$ no
tidal effects, $dd=$ tidal radius computed at the mean orbital radius,
$d5=$ tidal radius computed at the pericenter).\\

\begin{table}[h]
\begin{center}
\begin{tabular}{|c|c|c|c|c|c|c|}
\hline \hline
  & \multicolumn{6}{|c|}{$\delta \Phi$} \\ \hline
Galaxy & {\sl \tiny 0510dd}  & {\sl \tiny 0520dd} & {\sl \tiny 3010dd}& {\sl \tiny 3020dd} & {\sl \tiny 0520d0}& {\sl  \tiny 0520d5} \\ \hline \hline
LMC (Moore) & -0.12 & -0.21 & -0.12 & -0.21 & +1.35 & -0.22 \\ \hline
LMC (NFW) & -0.07 & -0.18 & -0.09 & -0.18 & +0.90 & -0.18  \\ \hline
M31 (Moore) &  +0.25 & +0.52 & +0.03 & +0.03 & +2.69 & -0.10 \\ \hline
M31 (NFW) & +0.78 & +1.59 & +0.02 & +0.05 & +1.93 & +2.09  \\ 
\hline \hline
\end{tabular}
\caption{\label{tab2} 
The effect of including sub-halos on the $\gamma$-ray flux from LMC and M31 
in different halo models. Each entry specifies the fractional flux
variation with respect to the smooth halo model.
The choice of parameters is encoded in each model label as specified in 
the text.
}
\end{center}
\end{table}

In all {\sl dd} models considered, 
the flux from M31 is increased by the presence of sub-halos
while the flux from LMC decreases.
This discrepancy reflects the different effect of the tidal field that,
as we have seen, disrupts all the sub-halos near the 
halo center and thus decreases the flux from the nearby LMC, while 
the flux from M31, which is produced 
in a broader region, receives a significant contribution from
the nearby, foreground sub-halos in M31.
Varying $m_{cl}$ in the allowed range does not change the fluxes
appreciably, while increasing the core radius of the sub-halo 
distribution decreases the M31 flux significantly but does not change 
significantly the flux received from the nearer LMC.
Finally, neglecting the tidal field ({\sl d0} model) has the rather
obvious effect of increasing the flux from both extragalactic objects.
Enhancing the tidal disruption effect ({\sl d5} model) generally 
decrease the annihilation flux, apart from the case of M31-NFW,  
in which the flux is enhanced by the combined effect
of the halo distance and the large DM concentration.

\section{V. Detectability}

In this section we explore 
the actual detectability of the $\gamma$-ray fluxes from both
the extragalactic sources and the Galactic foreground.

\noindent 
Satellite borne experiments, such as GLAST \cite{GLAST} , have very small
effective areas at energies $>250 \ GeV$ and thus are 
inefficient in revealing  high energy photons
from neutralino annihilation.
The need for larger collecting areas can only be met by those ground-based 
experiments designed to detect the products of generic atmospheric showers.  
These detectors were not originally meant to explore exotic physics, yet
recent advances in technology have allowed to reach unprecedented high
sensitivities and they can now be used to detect high energy photons from 
neutralino annihilation. \\

{\sl $\check{C}$erenkov detectors.}
Electrons generated in showers emit ${\rm \check{C}}$erenkov 
light during their passage through the atmosphere. It
reaches the ground in the form of a $\sim 100$ meters wide
light pool. 
${\rm \check{C}}$erenkov detectors have good angular and energetic resolution, 
operate with a duty cycle of about 10\% and 
have a very good hadronic background rejection. 
Their small angle of view ($\sim 5^{\circ}$) makes them suitable
for the observations of point sources like LG galaxies. 
Currently operating $\check{C}$erenkov
telescopes such WHIPPLE \cite{WHIPPLE} and HEGRA \cite{HEGRA}
do not have enough sensitivity to detect faint extragalactic sources. 
Planned Active ${\rm \check{C}}$erenkov Telescopes (ACT) 
such as HESS \cite{HESS}, 
MAGIC \cite{MAGIC} and VERITAS \cite{VERITAS} with their lower 
energy threshold ($\sim 50 \ GeV$) and greater sensitivities
make them suitable for our purposes.

{\sl Large field of view arrays.}
Unconventional air shower arrays at energies $<$ TeV
constitute viable alternative to  $\check{C}$erenkov telescopes.
Unlike standard air shower arrays, these detectors can 
reconstruct lower energy showers by dense sampling of the 
shower particles.
This can be achieved by operating the 
detector at very high altitude in order to approach the
maximum size development of low energy showers and by using a full 
coverage layer of counters.
Operating large field of view arrays like ARGO \cite{ARGO} and MILAGRO 
\cite{MILAGRO} have worse angular resolution and background rejection
than  ACT detectors, but are able to explore the whole sky with a 
duty cycle of about 100 \%. \\

\begin{figure} 
\includegraphics[height=7cm,width=8cm]{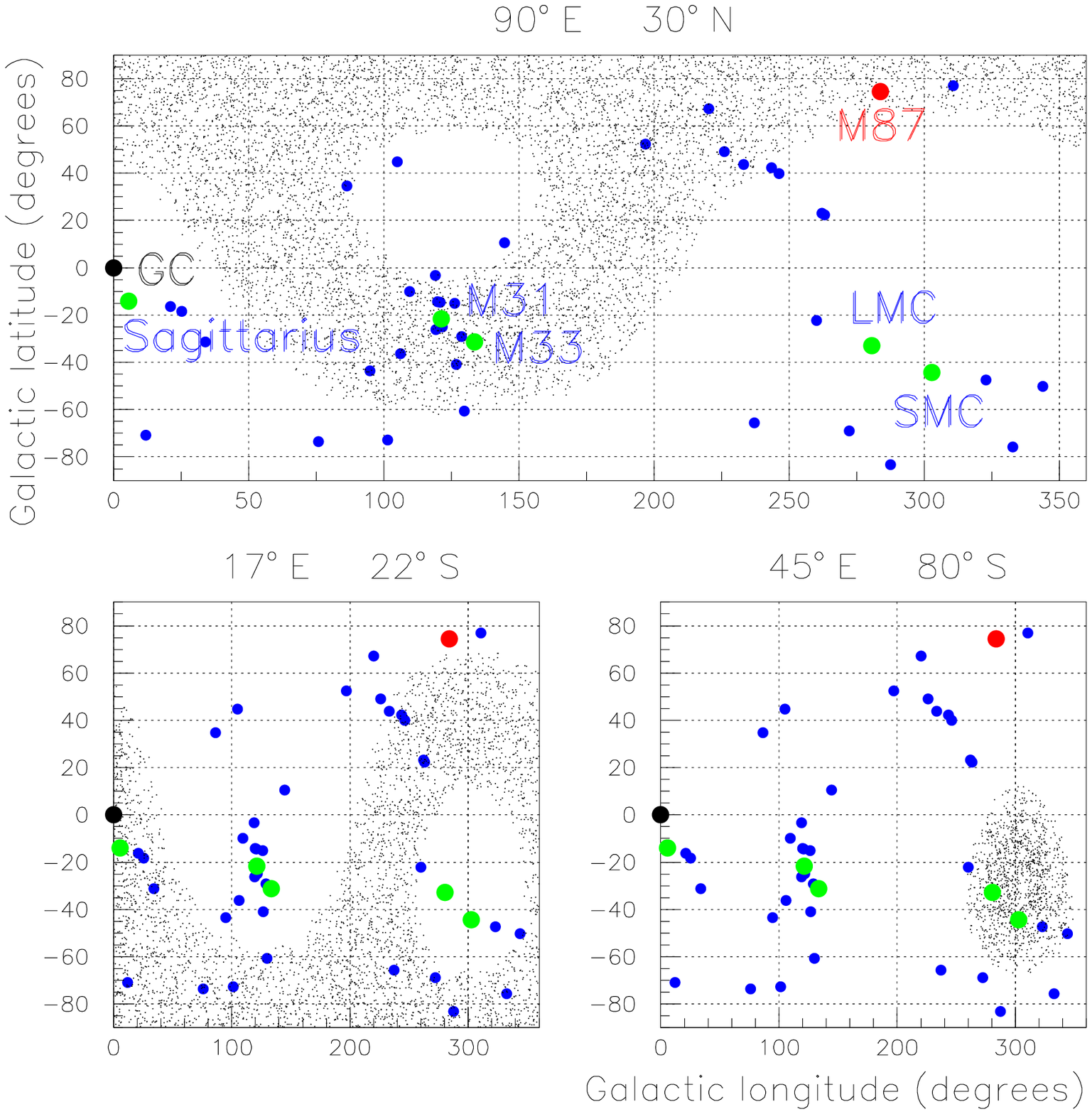}
\caption{Angular positions of our 44 LG galaxies (filled dots)
inclusive of M87 and the GC in galactic coordinates. 
Shaded regions  have zenith  angles smaller than $30^{\circ}$
at the ARGO site (upper panel),
HESS site (lower left) and an hypothetical Antarctic site
(lower right). Terrestrial longitude and latitude of each 
sites are indicated above each panel.
}
\label{fig6}
\end{figure}

Due to the complementarity of ACTs and large field of view arrays
we explore and compare the capabilities of these
detectors of revealing the $\gamma$-ray flux from neutralino
annihilation in extragalactic objects \cite{icrc2003}. 
All analyses, including our own, have shown that the best place
to look for DM annihilation signals is the GC, and indeed
presently available and future detectors should be able to
detect such $\gamma$-ray emission. 
However, most of them are (or will be)
located  in the Northern hemisphere where the GC can only be seen 
for short periods and at uncomfortably large zenith angles.
In Fig. \ref{fig6} the angular positions of our extragalactic objects
(filled dots) are shown in Galactic coordinates,
superimposed to the areas that can be seen  
 with a zenith angle smaller
than $30^{\circ}$ from various experimental sites (shaded areas).
In the following we estimate the detectability of  $\gamma$-ray fluxes
from either external LG galaxies or sub-halos within our Galaxy.

\subsection{Detectability of external galaxies}

In the following we will focus on the detectability of
M31 since it is the extragalactic object with the largest
$\gamma$-ray flux visible from the Northern hemisphere.

The detectability of the continuum flux from M31 is computed by comparing 
the number of $\gamma$ events expected from this source to the fluctuations
of background events. This ratio, which we call $\sigma$ in the following, 
is given by:
\begin{equation}
\frac{n_{\gamma}}{\sqrt{n_{bkg}}}= \frac{\sqrt{T_\delta} \epsilon_{\Delta \Omega}}{\sqrt{ \Delta \Omega}} \frac{\int A^{eff}_\gamma (E,\theta) \frac{\phi^{DM}_\gamma}{dE} dE d\theta}{\sqrt{ \int 
\sum_{bck} A^{eff}_{bck}(E,\theta) \frac{d\phi_{bck}}{dE} dE d\theta }},
\label{sensitivity}
\end{equation}
where $T_\delta$ is the time during which the source is seen with zenith angle
$\theta \leq 30^{\circ}$ and
$\epsilon_{\Delta \Omega} = 0.7$ is the fraction of signal events 
within the optimal solid angle $\Delta \Omega$ corresponding to the angular
resolution of the instrument.
The effective detection areas $A^{eff}$ for 
electromagnetic 
and hadronic induced showers are defined as the detection efficiency times
the geometrical detection area.
For the case of a large field of view array we have considered
$A^{eff}=10^2- 10^3 \ m^2$,  depending on the $\gamma$-photon 
energy, while for a 
 ${\rm \check{C}}$erenkov apparatus we have assumed
a conservative effective area of $10^4 \ m^2$.
Note that while the latter can be increased by adding
together  more ${\rm \check{C}}$erenkov telescopes,
the former is intrinsically limited by the 
number of hits reaching the ground
and cannot be much greater than our fiducial value.
Finally we have assumed an angular resolution
 $\phi= 1^{\circ}$ and hadron-photon identification 
efficiencies of $\epsilon=75 \%$ 
for the case of a large field of view array 
and $\phi= 0.1^{\circ}$  and $\epsilon=99 \%$ 
for the  ${\rm \check{C}}$erenkov apparatus.
These values are appropriate for the energy range of interest. \\

\begin{figure} 
\includegraphics[height=7cm,width=8cm]{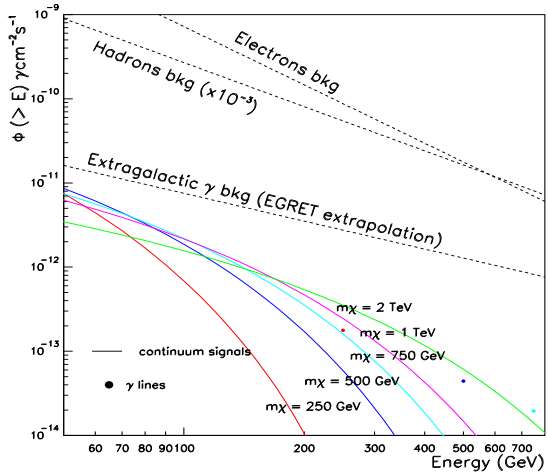}
\caption{Integrated spectra of photons from DM annihilation in M31
for different 
values of the neutralino mass, specified by the labels,
computed assuming a a Moore profile
filled with a population of sub-galactic halos unaffected by tides.
Both the continuum emission (curves) and the 
$\gamma$-line contribution (filled dots) are shown. The relevant backgrounds 
(dashed lines)
are shown for comparison.}
\label{fig7}
\end{figure}

{\sl Backgrounds.} In our analysis the M31 flux is 
much smaller than  electron,  hadron and diffuse $\gamma$-ray backgrounds
that must then be taken into account.
We have considered the following values for the background levels:
\begin{equation}
\frac{d \phi^h}{d\Omega dE} = 1.49 E^{-2.74} \frac{p}{cm^2 \ s \ sr \ GeV} 
\label{hadrons}
\end{equation}
for the hadronic background \cite{hbck},
\begin{equation}
\frac{d \phi^e}{d\Omega dE} = 6.9 \times 10^{-2} E^{-3.3} \frac{e}{cm^2 \ s \ sr \ GeV} 
\label{electrons}
\end{equation}
for the electron background \cite{elbck}, and
\begin{equation}
\frac{d \phi^\gamma_{diffuse}}{d\Omega dE} = 1.38 \times 10^{-6} E^{-2.1} 
\frac{\gamma}{cm^2 \ s \ sr \ GeV} 
\label{gammas}
\end{equation}
for the diffuse extragalactic gamma emission, 
as extrapolated from EGRET data at lower energies \cite{gbck}. \\
\noindent The Galactic diffuse emission along
M31 line of sight \cite{Bergstrom:98} turned out to be
negligible with respect to the annihilation signal and has been neglected.\\
In Fig. \ref{fig7} the various backgrounds (dashed lines)
are compared to the integrated energy spectra of $\gamma$-photons from
DM annihilation in M31, computed assuming a Moore DM profile 
and a population of sub-halos unaffected by tidal effects.
The various curves refer to different neutralino masses, as 
specified by the labels. 
Filled dots represent the expected $\gamma$-line fluxes. \\

\begin{figure} 
\includegraphics[height=7cm,width=8cm]{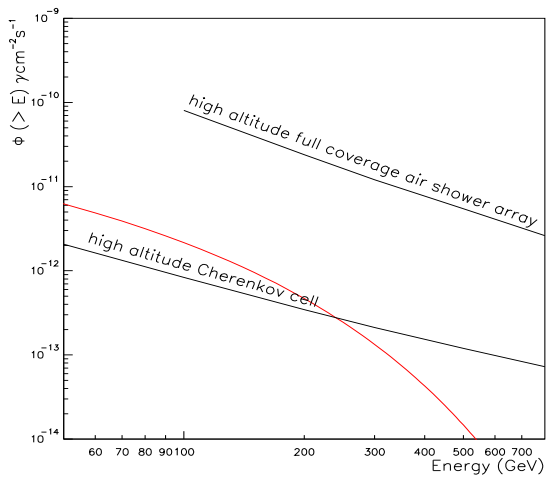}
\caption{$5 \sigma$ sensitivity curves for a high altitude full
coverage air shower array (upper line) and for a ${\rm \check{C}}$erenkov cell 
(lower line), computed using the parameters specified in the text.
The curve shows the same flux from M31 displayed in Fig. 6 for
the case of a neutralino mass of 1 Tev.}
\label{fig8}
\end{figure}

Fig. \ref{fig8} compares the expected integrated $\gamma$ flux from M31 
(dot-dashed curve) to the  $5 \ \sigma$ sensitivity 
curves for the  large field of view array (upper line) and the  
${\rm \check{C}}$erenkov telescope (lower line).
The  $5 \ \sigma$ detection curves assume
1 year of data taking for the large field of view array 
and 20 days pointing for the
${\rm \check{C}}$erenkov telescope. \\ 
The results show that annihilation signals from M31, which are too faint for  
large field of view arrays, are well within reach of 
next generation ${\rm \check{C}}$erenkov telescopes.

\subsection{Detectability of sub-halos in our Galaxy}

The only way of detecting signals from annihilation
in Galactic sub-halos is through blind searches
since model predictions specify the sub-halo distribution 
but not their precise spatial positions.
With this respect, the use of a large field of view array 
has the advantage of collecting signals from a wide area of the sky, 
while ACTs would only allow to survey a much smaller region.

To estimate the chances of detecting a $\gamma$-ray 
signal with either detectors we performed ideal
observations with both instruments and computed the significance
of an annihilation signal using Eq. (\ref{sensitivity}).

We restrict our  observations to
all directions with $|b| < 30^\circ$ if $|l| < 40^\circ$ 
around the GC and $|b| < 10^\circ$  elsewhere
provided that they are within a zenith angle of $30^\circ$. 
These constraints allow us to ignore 
the Galactic diffuse photon contribution
to the background emission and the effects of 
the atmospheric depth. 

\begin{figure} 
\includegraphics[height=7cm,width=8cm]{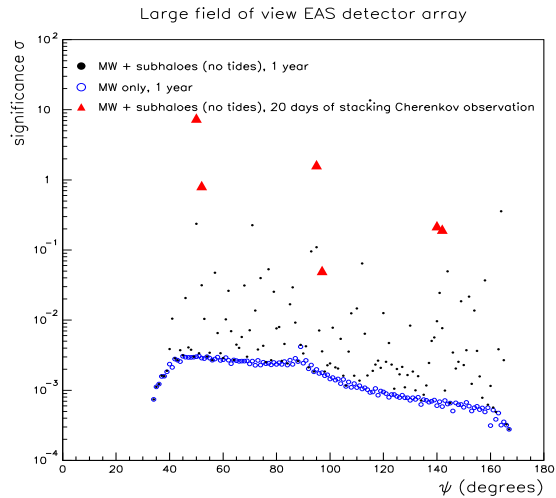}
\caption{Significance of the $\gamma$-ray signal from 
Galactic sub-halos as a function
of the angle of view from the GC, for a large field of view array 
(dots) and for a ${\rm \check{C}}$erenkov telescope (triangles).
The considered sub-halo distribution is not affected by any tidal effect.}
\label{fig9}
\end{figure}

For the case of a large field of view array
we have chosen the ARGO experimental site
($6.06 \ h$ E, $30.18^\circ$ N) in Tibet 
and considered the case of a 365 days observation
with a detector having an  
angular resolution of $1^{\circ}$.
Each point in  Fig. \ref{fig9} represents the 
significance of a $\gamma$-ray signal detection
in such experiment, plotted as a function of $\psi$,
the angle  from the GC.
Each filled dot accounts for the annihilation flux 
received from all Galactic sub-halos with the same 
$\psi$ that have entered  the detector field of view
during the observation. The Galactic signal in absence
of sub-halos is also plotted for reference (open dots).
The significance detection level is very small at all 
$\psi$ but in a couple of cases where a very nearby
sub-halo crosses the line of sight. The results show that
it would be impossible, using a large field of view array,
to identify the Galactic sub-halo origin of any detected 
$\gamma$-ray flux.\\

For the case of an ACT we have chosen the
VERITAS site ($8.87 \ h$ W, $33^\circ$ N) in Arizona.
In this case we have considered the results of
a 20 days ``stacking'' observation in which 
an instrument with an angular resolution of 0.1 degrees 
points toward different directions characterized by the same value of $\psi$.
The results of this experiment are represented by the
triangles  in Fig. \ref{fig9}. Since each  point represents 
a 20 days observations we have sampled only a few angles $\psi$.
Although the significance of the ACT detections is higher
than in the case of a  large field of view array, 
the serendipitous nature of the $\gamma$-ray sources combined
with the small extension of the surveyed area make it very
hard to detect the annihilation signal from Galactic sub-halos
using ${\rm \check{C}}$erenkov telescopes. \\

We then conclude that only satellite borne experiments
like GLAST will have the chance to detect 
$\gamma$-ray photons from Galactic sub-halos and to study 
their spatial distribution \cite{aloisio:02}.
  
\section{VI. Discussion and conclusions}

In this work we have computed the expected $\gamma$-ray flux from
DM annihilation in the nearest 
LG galaxies, under the assumption that the bulk of DM is constituted
by neutralinos.
Predicted fluxes depend on a number of assumptions  
like the shape of the DM halo density profiles, 
the possible existence and the distribution 
of a population of sub-galactic halos and the dynamical influence 
of SMBHs  located at the halo centers.
Uncertainties in these theoretical models that propagates into
flux predictions have been evaluated by
systematically exploring the
space of the currently accepted model parameters.

It turns out that, in all cases considered,  
the fluxes from the Small and the Large
Magellanic Clouds, M31 and M87 are well above the level
of the $\gamma$-ray annihilation Galactic foreground.
Uncertainties in the predicted Galactic emission
are quite large (they vary by a factor $\sim 100$)
and mainly result
from the poorly known modeling of the DM sub-structures within
our Galaxy. This scatter actually increases when 
considering more exotic possibilities that would 
further boost up the Galactic foreground 
like the existence of caustics in the DM distribution \cite{Zeldovich:84}
or the presence of a population of mini sub-halos allowed by the 
very small cutoff mass in the CDM spectra \cite{Berezinsky:03}.
Neutralino annihilation in sub-halos contributes less
to $\gamma$-ray flux from extragalactic sources, resulting 
in a much smaller (by a factor of $\sim 10$) model uncertainties 
and more robust flux predictions.

In section II we have assumed that the inner slope of the DM density
profile $\rho(r) \propto r^{-\alpha}$ is in the range
$1.0< \alpha < 1.5$ following the indications of
numerical experiments rather than using observational constraints.
The reason was that current observations do not provide a self-consistent
scenario.
Indeed, recent determinations based on the X-ray properties of
the intercluster medium seem to indicate a steep DM density
profile $1.0< \alpha < 2.0$, provided that clusters with disturbed
X-ray surface brightness are removed from the sample
\cite{buote:03,bautz:03}.
On the other hand, the rotation curves measured in a sample of LSBs by
\cite{deBlok:03} are consistent, on average, with a shallower 
density profile $\alpha = 0.2$.
However, when the full extent of the rotation curves is taken into
account rather than the inner region only, then the recovered
density profiles  lie between the NFW and the Moore ones
\cite{Hayashi:03}.
Another study of high resolution $H \alpha$ rotation curves for dwarf and
LSB \cite{swaters:03} has
ruled out neither a profiles as steep as $\alpha = 1$ nor a constant
density core.
Finally, indications for shallow profiles ($\alpha = 0.52 \pm 0.3$)
were found by combining stellar dynamics and strong lensing data in a sample
of brightest cluster galaxies \cite{sand:03},
and from microlensing events toward the GC
($\alpha = 0.4$, in the hypothesis that the MW halo is spherically
symmetric \cite{merrifield:03}).
However, estimates based on weak gravitational lensing in 
X-ray luminous clusters indicate a much steeper profile with
 $0.9 < \alpha < 1.6$ \cite{dahle:03}. \\
The previous examples show that the current observations neither support
nor reject profiles as shallow as $\alpha = 0.5$ or 0. Therefore, while
we regard the NFW and the Moore profiles as the most likely
cases based on N-body results, we also want to keep an eye on the
shallow profile case. For this purpose we have estimated the
expected $\gamma$-ray flux from both M31 and LMC in the case
of shallow profiles.
If $\alpha = 0.5$ then the M31 flux is $\sim 50$ time smaller
than in the NFW case and decreases by a further factor of 10 with a flat
core $\alpha = 0$. Results for the LMC are very similar.
Therefore, if the slope of the density profile would turn out to be
as shallow as $\alpha = 0.5$ or more, then no ground-based experiments
would have the chance of detecting extragalactic $\gamma$-ray annihilation
fluxes. \\

In the hypothesis of both a NFW and a Moore DM density profile
we have estimated the possibility of detecting the annihilation signals 
from Galactic sub-halos and extragalactic sources
using those ground based instruments sensitive enough 
to detect $\gamma$-photons
in the $\sim 50 \ GeV$ - few $TeV$ energy band, 
such as ACTs or large field of view arrays.
We have found that the expected fluxes from Galactic sub-halos  are too faint 
to be detected by ground-based observatories. Their existence and distribution
could instead be probed by next generation satellite borne experiments
such as GLAST.
On the other hand, ground-based experiments should be able to detect
a few extragalactic objects such as LMC and M31.
In particular, 
we have been focusing on M31 which turned out to be the brightest
extragalactic object visible in most experimental sites.
We have shown that 
next generation ${\rm \check{C}}$erenkov telescopes
should be capable of revealing the $\gamma$-ray flux from M31
at a significance level $\simeq 5 \sigma$ in a 20 days pointing observation. 
These results have been obtained using a quite conservative 
observational set up and theoretical modeling. Increasing the number of 
${\rm \check{C}}$erenkov cells (as already planned for some future experiments)
or the serendipitous superposition of a nearby Galactic sub-halo
would certainly increase the chance of 
detecting an annihilation signal along the M31 direction.
The capability of detecting extragalactic signals is not shared
by large shower arrays, which due to their lower sensitivity
will not be able to detect such a faint source. 
It is worth stressing that the GC does remain the best 
place to consider for detecting DM annihilation signatures.
Unfortunately, and unlike M31, it is not visible from the Northern 
hemisphere where most of the ground-based detectors are 
(or will be) located. 

Finally, it is worth stressing that our predictions assume that
DM is made of neutralinos.
However, the natural factorization of particle physics and 
cosmology in the problem addressed in this paper 
makes straightforward  to extend our approach to other kinds of 
weakly interacting DM candidates.

\section{VII. Acknowledgments}
We thank M. De Vincenzi, N. Fornengo, P. Gondolo, 
S. Matarrese and  B. Moore for useful discussions and suggestions,
and J. Navarro and E. Romano-Diaz for having provided numerical codes
and graphic tools used in this work.

\end{document}